%% file: qz3l.tex
\newcommand*{\ATLASLATEXPATH}{./}
\documentclass[cernpreprint,texlive=2011,txfonts,UKenglish,texmf]{\ATLASLATEXPATH atlasdoc}
\pdfoutput=1 
\usepackage[biblatex=true,backend=bibtex]{\ATLASLATEXPATH atlaspackage} 
\usepackage[articletitle=false]{\ATLASLATEXPATH atlasbiblatex}

\usepackage{\ATLASLATEXPATH atlascontribute}

\usepackage{slashed}
\usepackage{bm}

\usepackage{\ATLASLATEXPATH atlasphysics}

\newcommand*{\tqz}{\ensuremath{t\to qZ}}
\newcommand*{\ttv}{\ensuremath{t\bar tV}}
\newcommand*{\ttw}{\ensuremath{t\bar tW}}
\newcommand*{\ttz}{\ensuremath{t\bar tZ}}
\newcommand*{\tth}{\ensuremath{t\bar tH}}

\hypersetup{pdftitle={ATLAS draft},pdfauthor={The ATLAS Collaboration}}

\input{qz3l-metadata.tex}

\begin{document}

\maketitle


\input{intro.tex}
\input{detector.tex}
\input{samples.tex}
\input{reconstruction.tex}
\input{selection.tex}
\input{backgrounds.tex}
\input{systematics.tex}
\input{limits.tex}
\input{conclusions.tex}

\section*{Acknowledgements}
\input{Acknowledgements.tex}

\printbibliography

\newpage
\input{atlas_authlist}

\end{document}

%% file: qz3l-metadata.tex
\AtlasTitle{Search for flavour-changing neutral current top-quark decays to $\bm{qZ}$ in $\bm{pp}$ 
collision data collected with the ATLAS detector at $\bm{\sqrt{s}=8}$~TeV}

\author{The ATLAS Collaboration}

\AtlasRefCode{TOPQ-2014-08}

\PreprintIdNumber{CERN-PH-EP-2015-183}

\AtlasDate{\today}

\AtlasJournal{Eur.\ Phys.\ J.\ C}

\AtlasAbstract{%

A search for the flavour-changing neutral-cur\-rent decay $t\to qZ$ is 
presented. Data collected by the ATLAS detector during 2012 from 
proton--proton collisions at the Large Hadron Collider at a 
centre-of-mass energy of $\sqrt{s}=8$~TeV, corresponding to an integrated 
luminosity of 20.3~fb$^{-1}$, are analysed. Top-quark pair-production events 
with one top quark decaying through the $t\to qZ$ ($q=u,c$) channel and the 
other through the dominant Standard Model mode $t\to bW$ are considered as 
signal. Only the decays of the $Z$ boson to charged leptons and leptonic $W$ 
boson decays are used. No evidence for a signal is found and an observed 
(expected) upper limit on the $t\to qZ$ branching ratio of $7\times 10^{-4}$ 
($8\times 10^{-4}$) is set at the 95\% confidence level.
}

%% file: intro.tex
\section{Introduction\label{sec:intro}}

The top quark is the heaviest elementary particle known, with a mass 
$m_t=173.21\pm0.51({\mathrm{stat.}})\pm0.71({\mathrm{syst.}})$~GeV~\cite{Agashe:2014kda}. 
Its lifetime is so short that, within the Standard Model~(SM) of particle 
physics, it decays (almost exclusively to $bW$) before hadronisation 
occurs. These properties make it a particle well suited to test the 
predictions of the SM. In the SM, due to the GIM 
mechanism~\cite{Glashow:1970gm}, flavour-changing neutral current~(FCNC) 
decays such as $t\to qZ$ are forbidden at tree level. They are allowed at one-loop level, but with a suppression factor of
several orders of magnitude with respect to the dominant decay mode~\cite{AguilarSaavedra:2004wm}. However, several SM extensions predict higher branching 
ratios~(BRs) for the top-quark FCNC decays. Examples of such extensions are 
the quark-singlet model (QS)~\cite{AguilarSaavedra:2002kr}, the two-Higgs-doublet
model with~(FC 2HDM) or without~(2HDM) 
flavour conservation~\cite{Atwood:1996vj}, the minimal supersymmetric model 
(MSSM)~\cite{Cao:2007dk}, supersymmetry with R-parity 
violation~($\slashed{R}$~~SUSY)~\cite{Yang:1997dk} or models with warped 
extra dimensions~(RS)~\cite{Agashe:2006wa}. For a review see 
Ref.~\cite{Agashe:2013hma}. The maximum values for the \tqz\ BRs predicted by these 
models and by the SM are summarised in Table~\ref{tab:intro:models}. 
Experimental limits on the FCNC \tqz\ BR were established by experiments at 
the Large Electron Positron Collider~(LEP)~\cite{Heister:2002xv, 
Abdallah:2003wf, Abbiendi:2001wk, Achard:2002vv, LEP-Exotica-WG-2001-01}, 
HERA~\cite{Abramowicz:2011tv}, Tevatron~\cite{Aaltonen:2008ac,Abazov:2011qf} 
and Large Hadron Collider~(LHC)~\cite{TOPQ-2011-22, CMS-TOP-12-037}. The most 
stringent limit, BR$(t\to qZ) < 5\times 10^{-4}$, is the one from the CMS Collaboration~\cite{CMS-TOP-12-037} 
using 25~fb$^{-1}$ of data collected at $\sqrt{s}=7$~TeV and 
$\sqrt{s}=8$~TeV. Previous ATLAS results obtained at $\sqrt{s}=7$~TeV are 
also reported~\cite{TOPQ-2011-22}. Limits on other FCNC top-quark decay BRs 
($t\to qX, X=\gamma,g,H$) are reported in Refs.~\cite{Heister:2002xv, 
Abdallah:2003wf, Abbiendi:2001wk, Achard:2002vv, LEP-Exotica-WG-2001-01, 
Aaron:2009vv, Abe:1997fz, Aaltonen:2008qr, Abazov:2010qk, TOPQ-2014-13,
HIGG-2013-09, TOPQ-2014-14, CMS-HIG-13-025, CMS-TOP-14-003}.

This paper presents the ATLAS results from the search for the FCNC decay \tqz\ in 
\ttbar\ events produced at $\sqrt{s}=8$~TeV, with one top quark decaying 
through the FCNC mode and the other through the SM dominant mode ($t\to bW$). Only 
the decays of the $Z$ boson to charged leptons and leptonic $W$ boson decays 
are considered. The final-state topology is thus characterised by the 
presence of three isolated charged leptons, at least two jets, and missing 
transverse momentum from the undetected neutrino. The paper is organised as 
follows. A brief description of the ATLAS detector is given in Section 
\ref{sec:detector}. The collected data samples and the simulations of signal 
and SM background processes are described in Section 
\ref{sec:samples}. Section \ref{sec:reco} presents the object definitions, 
while the event analysis and kinematic reconstruction are explained in Section 
\ref{sec:selection}. Background evaluation and sources of systematic 
uncertainty are described in Sections \ref{sec:backgrounds} and 
\ref{sec:systematics}. Results are presented in Section \ref{sec:limits} and 
conclusions are drawn in Section \ref{sec:conclusions}.

%% file: detector.tex
\section{Detector and data samples\label{sec:detector}}

The ATLAS experiment is a multi-purpose particle physics detector consisting 
of several sub-detector systems, which cover almost fully the solid 
angle\footnote{\label{dr}ATLAS uses a right-handed coordinate system with its 
origin at the nominal interaction point in the centre of the detector and the 
$z$-axis along the beam pipe. The $x$-axis points from the interaction point 
to the centre of the LHC ring, and the $y$-axis points upward. Cylindrical 
coordinates $(r,\phi)$ are used in the transverse plane, $\phi$ being the 
azimuthal angle around the beam pipe. The pseudorapidity is defined in terms 
of the polar angle $\theta$ as $\eta=-\ln\tan(\theta/2)$. The $\Delta R$ 
distance is defined as $\Delta R=\sqrt{(\Delta\eta)^2+(\Delta\phi)^2}$.} around 
the interaction point. It is composed of an inner tracking system close to 
the interaction point and immersed in a 2~T axial magnetic field produced by 
a thin superconducting solenoid, a lead/liquid-argon (LAr) electromagnetic 
calorimeter, an iron/scintillator-tile hadronic calorimeter, copper/LAr 
hadronic endcap calorimeter and a muon spectrometer with three 
superconducting magnets, each one with eight toroid coils. The forward region 
is covered by additional LAr calorimeters with copper and tungsten absorbers. 
The combination of all these systems provides charged-particle momentum 
measurements, together with efficient and precise lepton and photon 
identification in the pseudorapidity range $|\eta| < 2.5$. Energy deposits 
over the full coverage of the calorimeters, $|\eta| < 4.9$ are used to 
reconstruct jets and missing transverse momentum (with magnitude $\MET$). A three-level 
trigger system is used to select interesting events. The Level-1 trigger is 
implemented in hardware and uses a subset of detector information to reduce 
the event rate to a design value of at most 75~kHz. This is followed by two 
software-based trigger levels which together reduce the event rate to less 
than 1~kHz. A detailed description of the ATLAS detector is provided in 
Ref.~\cite{PERF-2007-01}.

In this paper the full 2012 dataset from proton--proton ($pp$) collisions at 
$\sqrt{s}=8$~TeV is used. The analysed events were recorded by 
single-electron or single-muon triggers and fulfil standard data-quality 
requirements. Triggers with different transverse momentum thresholds are used 
to increase the overall efficiency. The triggers using a low transverse 
momentum~($\pt$) threshold ($\pt^{e,\mu}>24$~GeV) also have an isolation 
requirement. Efficiency losses at higher \pt\ values are recovered by higher 
threshold triggers ($\pt^e>60$~GeV or $\pt^\mu>36$~GeV) without any isolation 
requirement. The integrated luminosity of the analysed data sample is 20.3 
fb$^{-1}$.

\begin{table}[t]

\caption{Maximum allowed FCNC \tqz\ BRs as predicted by several 
models~\cite{AguilarSaavedra:2004wm, AguilarSaavedra:2002kr, Atwood:1996vj, 
Cao:2007dk, Yang:1997dk, Agashe:2006wa, Agashe:2013hma}}
\label{tab:intro:models}

\begin{center}
\begin{tabular}{lcccccccc}
  \hline
  Model:		& SM		& QS		& 2HDM		& FC 2HDM	& MSSM		& $\slashed{R}$~~SUSY	& RS			\\
  \hline\\[-1em]
  BR$(t\to qZ)$:	& $10^{-14}$	& $10^{-4}$	& $10^{-6}$	& $10^{-10}$	& $10^{-7}$	& $10^{-6}$		& $10^{-5\phantom{0}}$	\\
  \hline
\end{tabular}
\end{center}

\end{table}

%% file: samples.tex
\section{Simulated samples\label{sec:samples}}

In the SM, top quarks are produced at the LHC mainly in pairs, with a 
predicted $t\bar{t}$ cross section in $pp$ collisions at a centre-of-mass 
energy of $\sqrt{s} = 8$~\tev of $\sigma_{t\bar{t}}= 253^{+13}_{-15}$~pb for 
a top-quark mass of $172.5$~\gev. The cross section has been calculated at next-to-next-to 
leading-order (NNLO) in QCD including resummation of next-to-next-to leading 
logarithmic (NNLL) soft gluon terms with {\tt Top++}
2.0~\cite{Cacciari:2011hy,Baernreuther:2012ws,Czakon:2012zr,Czakon:2012pz,Czakon:2013goa,Czakon:2011xx}. 
The parton distribution function (PDF) and $\alpha_{\rm S}$ uncertainties are calculated using the PDF4LHC 
prescription~\cite{Botje:2011sn} with the MSTW 2008 68\% CL 
NNLO~\cite{Martin:2009iq,Martin:2009bu}, CT10 
NNLO~\cite{Lai:2010vv,Gao:2013xoa} and NNPDF 2.3 5f FFN~\cite{Ball:2012cx} 
PDF sets and are added in quadrature to the renormalisation and factorisation scale uncertainties. The cross-section value 
for the NNLO+NNLL calculation is about 3\% larger than the exact NNLO prediction
implemented in {\tt HATHOR} 1.5~\cite{Aliev:2010zk}.

The simulation of signal events is performed with {\tt PROTOS} 
2.2~\cite{AguilarSaavedra:protos, AguilarSaavedra:2010rx}, which includes the 
effects of new physics at an energy scale $\Lambda$ by adding dimension-six 
effective terms to the SM Lagrangian. The most general $Ztu$ vertex that 
arises from the dimension-six operators can be parameterised including only 
$\gamma^\mu$ and $\sigma^{\mu\nu}q_{\nu}$ 
terms~\cite{AguilarSaavedra:2008zc} as:
\begin{eqnarray}
\mathcal{L}_{Ztu} & = & -\frac{g}{2c_W}\bar u\gamma^{\mu}\left(X^{\rm L}_{ut}P_{\rm L}+X^{\rm R}_{ut}P_{\rm R}\right)tZ_{\mu}\nonumber\\
& & 		  -\frac{g}{2c_W}\bar u\frac{i\sigma^{\mu\nu}q_{\nu}}{m_Z}\left(\kappa^{\rm L}_{ut}P_{\rm L}+\kappa^{\rm R}_{ut}P_{\rm R}\right)tZ_{\mu} + \mathrm{h.c.},
\end{eqnarray}
where
$g$ is the electroweak coupling,
$c_W$ is the cosine of the weak mixing angle,
$u$ and $t$ are the quark spinors,
$Z_\mu$ is the $Z$ boson field,
$P_{\mathrm{L}}$ ($P_{\mathrm{R}}$) is the left-handed (right-handed) projection operator,
$m_Z$ is the $Z$ boson mass and
$q^{\nu}=p^{\nu}_t-p^{\nu}_u$ is the outgoing $Z$ boson momentum.
The $Ztc$ vertex can be parameterised
in a similar fashion. This vertex involves a minimum of four anomalous couplings 
$X^{\mathrm{L}}_{ut}, X^{\mathrm{R}}_{ut}, \kappa^L_{ut},\kappa^R_{ut}$, which are set to 0.01 
each. It was checked that the coupling choice does not affect the kinematics 
of the event. No impact in the kinematics is seen by comparing the $bWuZ$ and 
$bWcZ$ processes and the latter is used as reference. Only decays of the $W$ and $Z$ bosons involving charged 
leptons are generated at the matrix-element level by {\tt PROTOS} ($Z\to 
e^+e^-,\mu^+\mu^-,\tau^+\tau^-$ and $W\to e\nu,\mu\nu,\tau\nu$). The 
CTEQ6L1~\cite{Pumplin:2002vw} leading-order PDF is used. To account for 
higher-order contributions in the signal production, the events are 
reweighted to the measured $t\bar t$ differential cross section as 
a function of the transverse momentum of the $t\bar t$ system 
$(1/\sigma)(\mathrm{d}\sigma/\mathrm{d}p_{\mathrm{T}}^{t\bar 
t})$~\cite{TOPQ-2012-08}. 
Hadronisation is handled by {\tt PYTHIA} 
6.426~\cite{Sjostrand:2006za} with the Perugia2011C~\cite{Skands:2010ak} set of tuned parameters 
and $\tau$ decays are processed with {\tt TAUOLA}~\cite{Jadach:1990mz}. 
The top-quark mass is set to $m_t = 172.5$~GeV. Additional simulations 
with different parton shower parameterisations are used to estimate the 
systematic uncertainties on the amount of initial- and final-state radiation~(ISR/FSR).

Several SM processes have final-state topologies similar to the signal, with 
at least three prompt charged leptons, especially $WZ$, $ZZ$, 
\ttv,\footnote{$V$ stands for $W$ and $Z$ bosons. The simulation of \ttw\ 
events also includes a possible second $W$ boson.} \tth, $ggH$, $VH$, $tZ$ and 
triboson ($WWW$, $ZWW$ and $ZZZ$) production. Events with non-prompt leptons 
or in which at least one jet 
is misidentified as an isolated charged lepton (labelled as “fake 
leptons” throughout this paper) can also fulfil the event selection 
requirements. These events, typically $Z+$jets, $Z+\gamma$, \ttbar and 
single-top, are estimated from a data-driven method using a parameterisation 
of the true- and fake-lepton efficiencies. Samples of simulated events of 
these backgrounds with fake leptons are used to cross-check the data-driven 
estimation. The $Z+$jets simulations include $Z$ production in association 
with heavy-flavour quarks.

Table~\ref{tab:samples} summarises the information about the generators, 
parton shower and parton distribution functions used to simulate the 
different event samples considered in the analysis. Diboson events ($WZ$ and 
$ZZ$, where $Z$ means $Z/\gamma^*$) produced using {\tt SHERPA} contain up to 
three additional partons and are selected to have leptons with $\pt > 5$~GeV and 
$m_{\ell\ell} > 0.1$~GeV for the $Z/\gamma^*$. The additional $WZ$ and $ZZ$ 
samples are used for comparison. The $WZ$ {\tt ALPGEN} samples are simulated 
with up to five additional partons from the matrix element. The $ZZ$ {\tt 
HERWIG}~\cite{Corcella:2002jc} samples are selected to have one lepton with $\pt 
> 10$~GeV and $|\eta| < 2.8$. The simulations of \ttz, \ttw$(W)$, $tZ$ and 
tribosons include events with up to two extra partons in the final state. The simulated 
samples used to cross-check the data-driven estimation of background with fake 
leptons are also listed in Table~\ref{tab:samples}.

Detailed and fast simulations of the detector and trigger are performed with standard 
ATLAS software using {\tt GEANT4}~\cite{Agostinelli:2002hh, SOFT-2010-01} and 
{\tt ATLFASTII}~\cite{SOFT-2010-01}, respectively. The same offline 
reconstruction methods used on data are applied to the simulated samples.

\begin{table}

\caption{Generators, parton shower, parton distribution functions and 
parameter tune for hadronisation used to produce simulated samples used in this analysis.}

\label{tab:samples}

\begin{center}
\footnotesize
\begin{tabular}{lllll}
\hline
Sample						& Generator					& Parton shower					& PDF		& Tune		\\
\hline
\multicolumn{5}{l}{Samples with at least three prompt leptons:}														\\
$\ttbar\to bWqZ$				& {\tt PROTOS} 2.2~\cite{AguilarSaavedra:protos}& {\tt PYTHIA} 6.426~\cite{Sjostrand:2006za}	& CTEQ6L1~\cite{Pumplin:2002vw}	& Perugia2011C~\cite{Skands:2010ak}	\\
$WZ$						& {\tt SHERPA} 1.4.3~\cite{Gleisberg:2008ta}	& {\tt SHERPA} 1.4.3				& CT10~\cite{Lai:2010vv}	& ---		\\
$WZ$						& {\tt ALPGEN} 2.14~\cite{Mangano:2002ea}	& {\tt HERWIG} 6.520.2~\cite{Corcella:2002jc}			& CTEQ6L1			& AUET2~\cite{ATL-PHYS-PUB-2011-008}	\\
$ZZ$						& {\tt SHERPA} 1.4.3				& {\tt SHERPA} 1.4.3				& CT10				& ---					\\
$ZZ$						& {\tt HERWIG} 6.5				& {\tt HERWIG} 6.5				& CTEQ6L1			& AUET2					\\
$\ttbar V$, $tZ$, tribosons			& {\tt MADGRAPH} 5 1.3.33~\cite{Alwall:2011uj}	& {\tt PYTHIA} 6.426				& CTEQ6L1			& AUET2B				\\
$\ttbar H$, $WH$, $ZH$				& {\tt PYTHIA} 8.163~\cite{Sjostrand:2014zea}	& {\tt PYTHIA} 8.163				& CTEQ6L1			& AU2~\cite{ATL-PHYS-PUB-2012-003}	\\
$ggH$						& {\tt POWHEG} 2~\cite{Frixione:2007vw}		& {\tt PYTHIA} 8.163				& CT10				& AU2					\\
\hline
\multicolumn{5}{l}{Other samples:}																					\\
$WW$						& {\tt SHERPA} 1.4.3				& {\tt SHERPA} 1.4.3				& CT10				& ---					\\
$Z+$jets (30~GeV~$ < m_{\ell\ell} < 1 $~TeV)	& {\tt ALPGEN} 2.14				& {\tt PYTHIA} 6.426				& CTEQ6L1			& Perugia2011C	\\
$Z+$jets (10~GeV~$ < m_{\ell\ell} < 60$~GeV)	& {\tt ALPGEN} 2.14				& {\tt HERWIG} 6.520.2				& CTEQ6L1			& AUET2		\\
$Z\gamma$					& {\tt SHERPA} 1.4.1				& {\tt SHERPA} 1.4.1				& CT10				& ---		\\
$\ttbar\to bWbW$				& {\tt POWHEG} 2				& {\tt PYTHIA} 6.426				& CTEQ6L1			& Perugia2011C	\\
Single top ($s$, $Wt$ channel)			& {\tt MC@NLO} 4.03~\cite{Frixione:2002ik}	& {\tt HERWIG} 6.520.2				& CT10				& AUET2		\\
Single top ($t$ channel)			& {\tt AcerMC} 3.8~\cite{Kersevan:2004yg}	& {\tt PYTHIA} 6.426				& CTEQ6L1			& AUET2B	\\
\hline
\end{tabular}
\end{center}

\end{table}

%% file: reconstruction.tex
\section{Object reconstruction\label{sec:reco}}

The primary physics objects considered in this search are electrons, muons, 
\MET, jets, and $b$-tagged jets. Tau leptons are not explicitly 
reconstructed, although the $\tau$ decay products are reconstructed as 
electrons, muons or jets and as an additional contribution to the missing 
transverse momentum.

Electron candidates are reconstructed~\cite{PERF-2013-05} from energy 
deposits (clusters) in the electromagnetic calorimeter, which are then 
matched to reconstructed charged-particle tracks in the inner detector. 
The candidates are required to have a transverse energy \et greater than 
15~GeV and a pseudorapidity of the calorimeter cluster associated with the 
electron candidate $|\eta_{\mathrm{cluster}}|<2.47$. Candidates in the 
transition region between the barrel and endcap calorimeters with $1.37 
< |\eta_{\mathrm{cluster}}| < 1.52$ are excluded. Electron candidates in this 
analysis must satisfy tight quality requirements on the electromagnetic 
cluster and associated track which provide discrimination between isolated 
electrons and jets. In order to suppress multi-jet backgrounds, it is also 
required that there is little activity in the space surrounding the electron. 
Two isolation variables are employed: the energy deposited around the 
electron in the calorimeter in a cone of size $\Delta R = 0.2$ and the scalar sum of the \pt\ of 
the tracks within cone of size $\Delta R = 0.3$ around the electron. Cuts on these two 
quantities are used to select isolated electrons; the adopted cuts yield a 90\% 
identification efficiency in $Z$ boson decays to $e^+e^-$ events from the full 
2012 dataset. Additionally, the longitudinal impact parameter $|z_0|$ of the 
electron track with respect to the selected primary vertex of the event is 
required to be less than 2~mm. The closest jet if separated by $\Delta R < 0.2$ from the 
selected electron is removed from the event. The electron candidate is 
discarded if an additional selected jet is found with $\Delta R < 0.4$. A 
looser electron selection, used for the estimation of backgrounds with fake 
leptons, is defined by removing the isolation requirements.

The muon candidate reconstruction~\cite{PERF-2011-01} is performed by 
finding, combining and fitting track segments in the layers of the muon 
chambers, starting from the outermost layer. The identified muons are then 
matched with tracks reconstructed in the inner detector. The candidates are 
refitted using the complete track information from both detector systems, and 
are required to satisfy $\pt > 15$~GeV, $|\eta|<2.5$ and to be separated by 
$\Delta R > 0.4$ from any selected jet. The hit pattern in the inner detector 
is required to be consistent with a well-reconstructed track and the $|z_0|$ of 
the muon track is required to be less than 2 mm. Additionally, the sum of the 
momenta of tracks inside a cone around the muon candidate, with variable size 
such that it is smaller for higher muon \pt~\cite{Rehermann:2010vq}, must be less than 5\% of the 
muon energy. For the estimation of backgrounds with 
fake leptons, a looser selection is applied by removing the isolation 
requirement.

Jets are reconstructed~\cite{PERF-2012-01} from topological clusters of neighbouring calorimeter 
cells with significant energy deposits using the anti-{\it 
k$_t$} algorithm~\cite{Cacciari:2008gp} with a radius parameter $R=0.4$. 
Prior to jet finding, a local calibration scheme is applied to correct the 
topological cluster energies for the non-compensating response of the calorimeter, dead 
material and energy leakage. The corrections are obtained from simulations of 
charged and neutral particles. These jets are then calibrated to the hadronic 
energy scale using $\pt$- and $\eta$-dependent correction factors. Dedicated 
requirements are applied to remove the negligible fraction of events (less 
than 0.01\%) where a jet is incorrectly reconstructed from a few noisy 
calorimeter cells~\cite{ATLAS-CONF-2010-038}. The jets 
used in the analysis are required to have $\pt>25$~GeV and $|\eta|<2.5$. To 
reduce the number of selected jets that originate from secondary $pp$ 
interactions, for jets with $\pt < 50$~GeV and $|\eta|<2.4$, the scalar sum 
of the $\pt$ of tracks matched to a jet and originating from the primary 
vertex must be at least 50\% of the scalar sum of the $\pt$ of all tracks 
matched to the jet.

Jets containing $b$-hadrons are identified (`$b$-tagged')~\cite{ATLAS-CONF-2011-102} using an algorithm
based on multivariate techniques. It combines information from the impact 
parameters of displaced tracks and from topological properties of secondary 
and tertiary decay vertices reconstructed within the 
jet. It is determined with simulated $t\bar t$ 
events that, for the chosen working point, the tagging efficiency for $b$-jets 
with $\pt>20$~GeV is 70\%, while the rejection factors for light-quark or gluon jets (light jets), charm 
jets and $\tau$ leptons are 137, 5 and 13, respectively.

The measurement of \MET is based~\cite{PERF-2011-07} on the energy deposits 
in the calorimeter with $|\eta|<4.9$. The energy deposits associated with 
reconstructed jets and electrons are calibrated accordingly. Energy deposits 
not associated with a reconstructed object are calibrated according to their 
energy sharing between the electromagnetic and hadronic calorimeters. The 
momentum associated with each reconstructed muon, estimated using the momentum 
measurement of its reconstructed track, is taken into account in the 
calculation of \MET.

%% file: selection.tex
\section{Event selection and kinematics\label{sec:selection}}

At least one of the selected leptons must be matched, with $\Delta R < 0.15$ to 
the appropriate trigger object and have $\pt > 25$~GeV. The trigger 
efficiencies for the leptons are approximately 93\% for electrons, 70\% for 
muons with $|\eta| < 1.05$ and 86\% for muons with $1.05 < |\eta| < 
2.4$~\cite{ATLAS-CONF-2012-048,TRIG-2012-03}. The events are required to have at least one 
primary vertex with more than four associated tracks, each with 
$\pt>400$~MeV. The primary vertex is chosen as the one with the highest $\sum \pt^2$ over all
associated tracks. Leptons from cosmic rays are rejected by removing
muon pairs with large, oppositely signed transverse impact parameters 
($|d_0|>0.5$~mm) and consistent with being back-to-back in the $r-\phi$ plane.
Events with noise 
bursts and readout errors in the LAr calorimeter are also rejected. Exactly 
three isolated leptons associated with the same vertex are required. The three 
leptons must have $|\eta|<2.5$ and $\pt > 15$~GeV. Two of the leptons are 
required to have the same flavour, opposite charge and a reconstructed 
mass within 15~GeV of the $Z$ boson mass 
($m_Z$)~\cite{Agashe:2014kda}. If more than one compatible lepton-pair is 
found, the one with the reconstructed mass closest to $m_Z$ is chosen as the 
$Z$ boson candidate. According to the signal topology, the events are then 
required to have $\MET>20$~GeV and two jets, although an additional third jet 
from initial- or final-state radiation is allowed. All jets are required to 
have $\pt > 35$~GeV and $|\eta|<2.5$. One or two of the jets must be 
$b$-tagged. Only one $b$-tagged jet is expected in the signal events, 
nevertheless a second one can arise from a misidentified $c$-jet associated with
the FCNC decay of the top quark. Allowing for the additional $b$-tagged jet increases 
the signal efficiency without compromising the signal-to-background ratio.

Applying energy--momentum conservation, the kinematics of the top quarks can 
be reconstructed from the corresponding decay particles. Since the neutrino 
from the semileptonic decay of the top quark ($t\to bW\to b\ell\nu$) is 
undetected, its four-momentum must be estimated. This can be done by 
assuming that the lepton not previously assigned to the $Z$ boson and the 
$b$-tagged jet (labelled $b$-jet) originate from the $W$ boson and SM top-quark decays,
respectively, and that \MET\ is the neutrino's
transverse momentum. The longitudinal component of the neutrino's momentum 
($p^{\nu}_z$) is then determined by minimising, without constraints, the 
following expression:
\begin{equation}
\chi^2  =  \frac{\left(m^{\mathrm{reco}}_{j_a\ell_a\ell_b}-m_{t_{\mathrm{FCNC}}}\right)^2}{\sigma_{t_{\mathrm{FCNC}}}^2}+
\frac{\left(m^{\mathrm{reco}}_{j_b\ell_c\nu}-m_{t_{\mathrm{SM}}}\right)^2}{\sigma_{t_{\mathrm{SM}}}^2}
 + \frac{\left(m^{\mathrm{reco}}_{\ell_c\nu}-m_W\right)^2}{\sigma_W^2},
\label{eq:chi2}
\end{equation}
where $m^{\mathrm{reco}}_{j_a\ell_a\ell_b}$, 
$m^{\mathrm{reco}}_{j_b\ell_c\nu}$ and $m^{\mathrm{reco}}_{\ell_c\nu}$ are 
the reconstructed masses of the $qZ$, $bW$ and $\ell\nu$ systems, 
respectively. The central value for the masses and the widths of the 
top quarks and $W$ boson are taken from reconstructed simulated signal 
events. This is done by matching the true particles in the simulated events 
to the reconstructed ones, setting the longitudinal momentum of the 
neutrino to the $p_z$ of the true simulated neutrino and then performing 
Bukin fits\footnote{This is a piecewise function with a Gaussian function in 
the centre and two asymmetric tails.}~\cite{Bukin} to the masses of the 
matched reconstructed top quarks and $W$ boson. The values are 
$m_{t_{\mathrm{FCNC}}}=173$~GeV, $\sigma_{t_{\mathrm{FCNC}}}=10$~GeV, 
$m_{t_{\mathrm{SM}}}=168$~GeV, $\sigma_{t_{\mathrm{SM}}}=23$~GeV, 
$m_W=82$~GeV and $\sigma_W=15$~GeV.

\begin{figure}
\begin{center}
  \includegraphics[width=.32\textwidth]{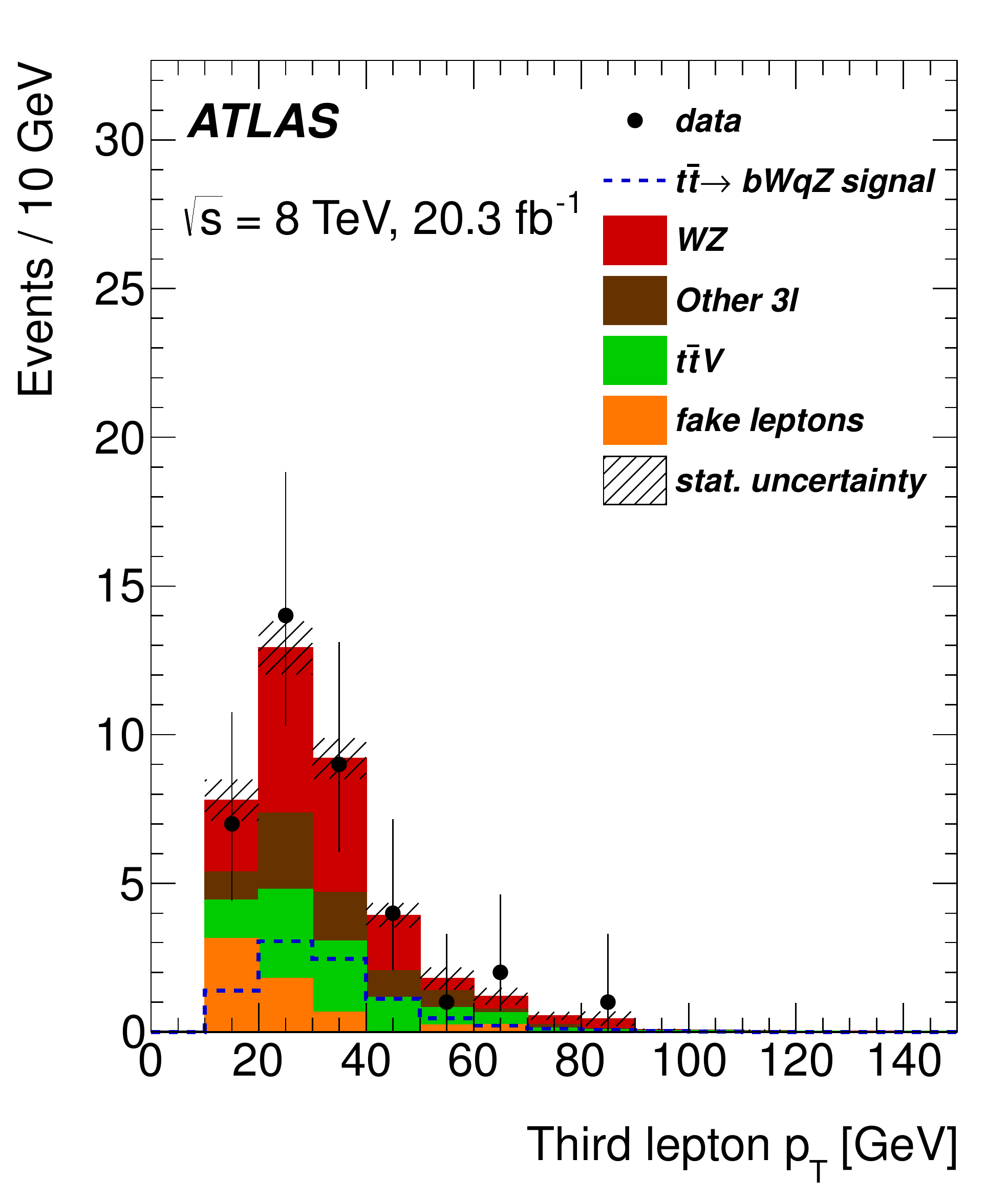}
  \includegraphics[width=.32\textwidth]{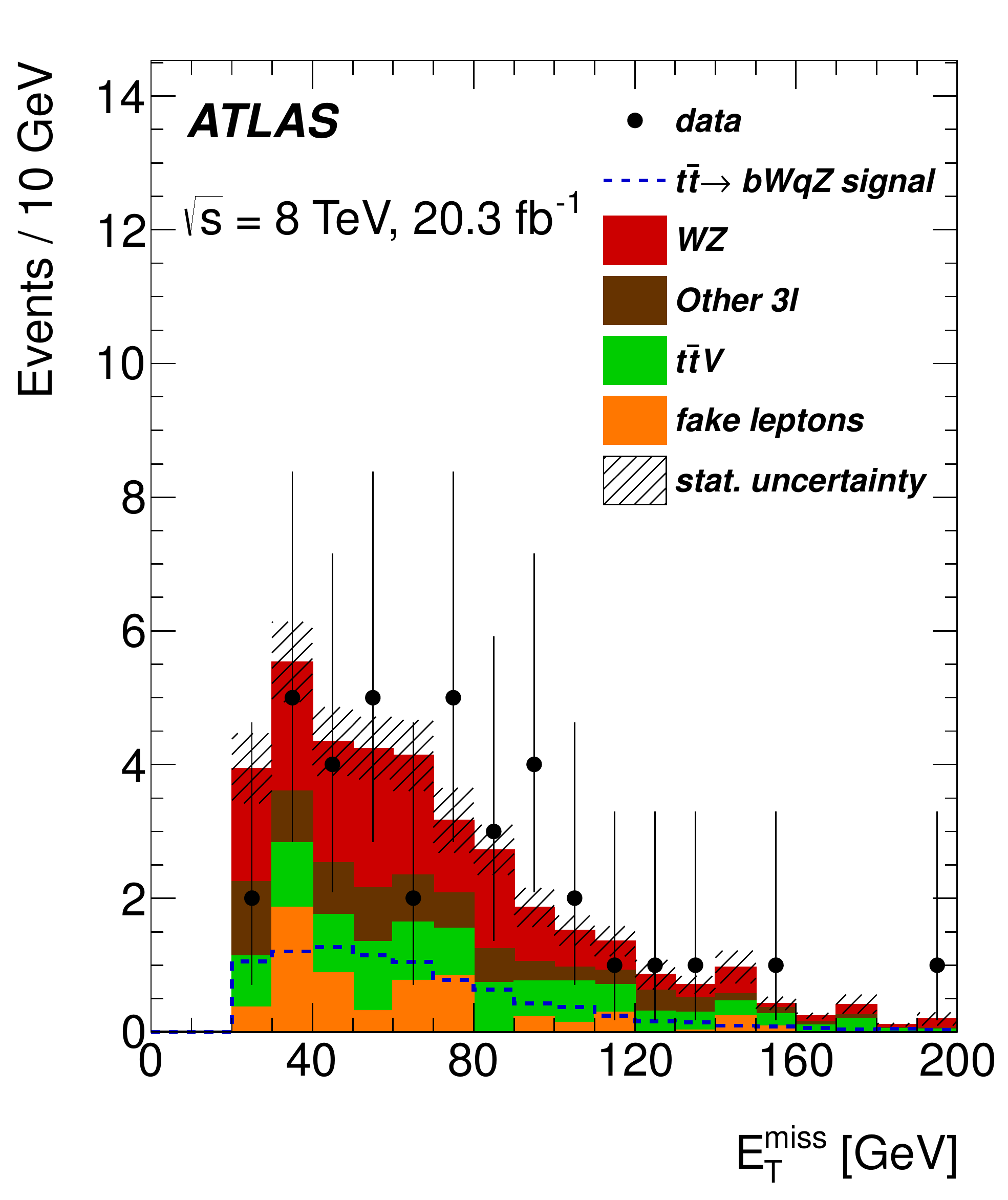}
  \includegraphics[width=.32\textwidth]{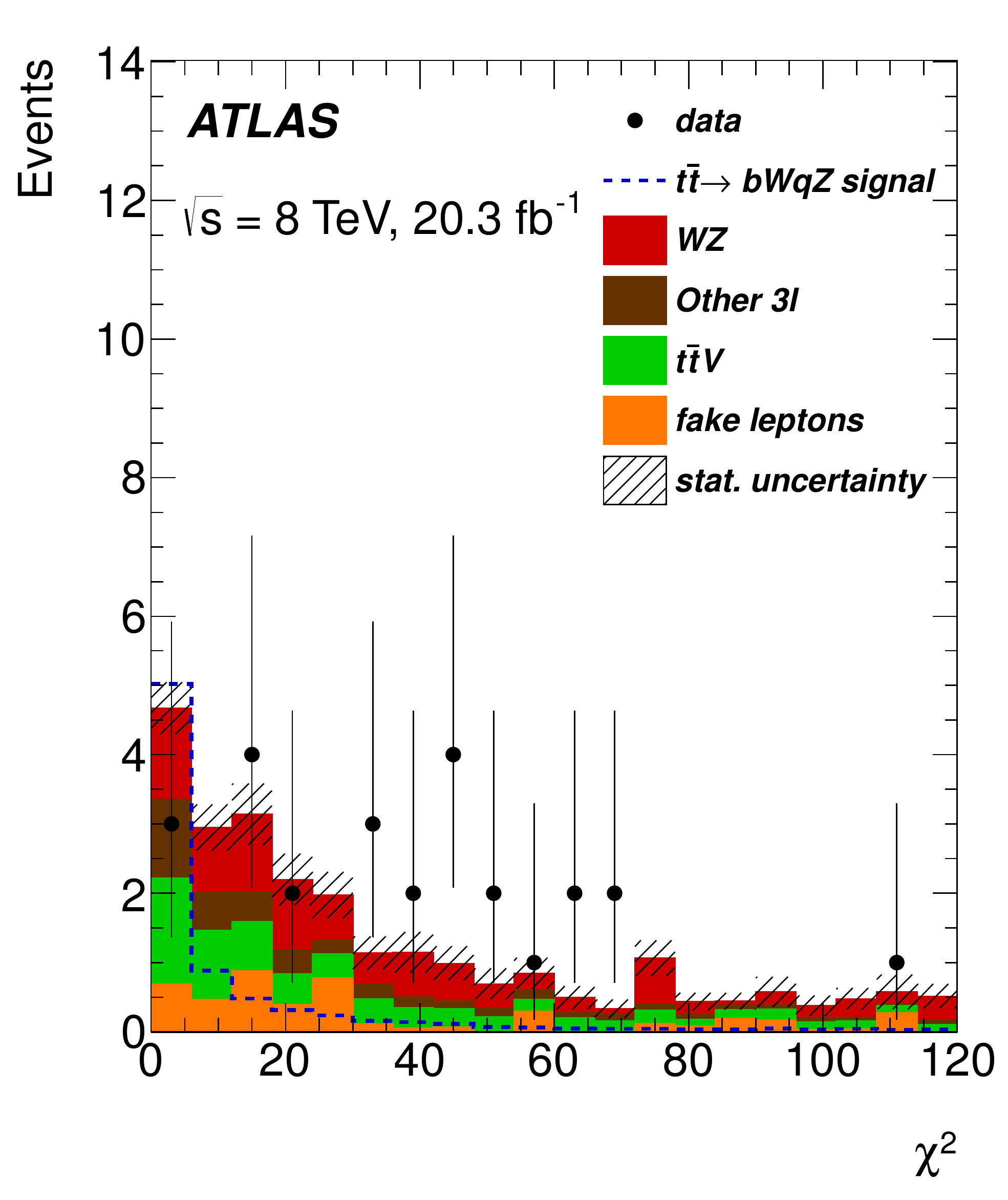}

  (a)\hspace*{.31\textwidth}(b)\hspace*{.31\textwidth}(c)
\end{center}

  \caption{Expected (filled histogram) and observed (points with error bars) distributions in the signal region before the $\chi^2$ cut is applied for
  (a) \pt\ of the third lepton,
  (b) \MET\ and
  (c) $\chi^2$ of the kinematics reconstruction.
  For comparison, distributions for the FCNC $\ttbar \to bWqZ$ signal (dashed 
  line), normalised to the observed 95\% CL limit reported in this paper, are also shown.
  Background statistical uncertainties associated with the number of events in the samples are represented by the hatched areas.
}

  \label{fig:preselection}
\end{figure}

For each jet combination, where any jet can be assigned to $j_a$, while $j_b$ 
must correspond to a $b$-tagged jet, the $\chi^2$ minimisation gives the most 
probable value for $p^{\nu}_z$. From all combinations, the one with the 
minimum $\chi^2$ is chosen, along with the corresponding $p^{\nu}_z$ value. 
The jet from the top-quark FCNC decay is referred to as the light-quark ($q$) 
jet. The fractions of correct assignments between the reconstructed top 
quarks and the true simulated particles (evaluated as a match within a cone of size
$\Delta R = 0.4$) are $\epsilon_{t_{\mathrm{FCNC}}}=79.9$\% and 
$\epsilon_{t_{\mathrm{SM}}}=56.3$\%.
Figure~\ref{fig:preselection} shows the 
\pt\ of the third lepton as well as the \MET\ and $\chi^2$ distributions at 
this level of the analysis.

The selection of the signal region is concluded with the requirement of 
$\chi^2<6$, which optimises the sensitivity discussed in 
Section~\ref{sec:limits}.

%% file: backgrounds.tex
\section{Background estimates\label{sec:backgrounds}}

Three control regions are defined to check the agreement between data and 
simulated samples of the $ZZ$, $WZ$ and \ttz\ backgrounds. No scaling factors 
are derived from these control regions, however they are used to estimate the 
background modelling uncertainties described in 
Section~\ref{sec:systematics}. The $tZ$ contribution to the total background 
is expected to be smaller than the one from \ttz\ 
events~\cite{Campbell:2013yla}. Due to the similarity between the final 
states of $tZ$ and signal events, there are large signal contributions to 
possible $tZ$ control regions.
For these reasons no control region is defined for the $tZ$ background.

The $ZZ$ control region is defined by requiring two pairs of leptons with the
same flavour, opposite charge and a reconstructed mass within 15~GeV of the $Z$ boson mass. The expected and observed yields are 
shown in Table~\ref{tab:backgrounds:zz_yields} and the {\tt SHERPA} sample is 
chosen as reference.

\begin{table}

\caption{Event yields in the $ZZ$ control region for all significant sources 
of background. The $ZZ$ {\tt SHERPA} sample is taken as reference for the 
total background estimation. The first uncertainty is the statistical one 
associated with the number of events in the simulated samples, the second 
uncertainty is systematic and is described in Section~\ref{sec:systematics}. The entry labelled ``other backgrounds'' includes all
the remaining backgrounds described in Section~\ref{sec:samples} and in 
Table~\ref{tab:samples}. The signal efficiency is also shown.}

\label{tab:backgrounds:zz_yields}

\begin{center}
\begin{tabular}{lc}
\hline
Sample			& Yields \\
\hline
$ZZ$ ({\tt SHERPA})	& $87\pm4\pm5$ \\
$ZZ$ ({\tt HERWIG})	& $85\pm4\pm5$ \\
Other backgrounds	& $0.48\pm0.05\pm0.08$\\
\hline
Total background	& $88\pm4\pm5$ \\
\hline
Data			& 95 \\
\hline
Signal efficiency $[\times 10^{-8}]$	& $5.6\pm4.3\pm0.1$ \\
\hline
\end{tabular}
\end{center}

\end{table}

To study the $WZ$ background the following control region is defined. Events 
are required to have three leptons, two of them with the same flavour, 
opposite charge and a reconstructed mass within 15~GeV 
of the $Z$ boson mass. Additional requirements include the presence of at 
least one jet in the event with $\pt > 35$~GeV, no $b$-tagged jets with $\pt 
> 35$~GeV and a $W$ boson transverse mass, built with the residual lepton and \MET, greater than 50~GeV. 
Table~\ref{tab:backgrounds:wz_yields} shows the expected and observed yields 
in this control region. The best estimation comes from the {\tt SHERPA} 
prediction, which is chosen as the reference sample.

\begin{table}

\caption{Event yields in the $WZ$ control region for all significant sources 
of background. The $WZ$ {\tt SHERPA} sample is taken as reference for the 
total background estimation. The first uncertainty is the statistical one 
associated with the number of events in the samples, the second 
uncertainty is systematic and is described in Section~\ref{sec:systematics}. The entry labelled 
``other backgrounds'' includes all the remaining
backgrounds described in Section~\ref{sec:samples} and in 
Table~\ref{tab:samples}. The signal efficiency
is also shown.}

\label{tab:backgrounds:wz_yields}

\begin{center}
\begin{tabular}{lc}
\hline
Sample			& Yields \\
\hline
$WZ$ ({\tt SHERPA})	& $333\pm5\pm17$ \\
$WZ$ ({\tt ALPGEN})	& $393\pm6\pm19$ \\

$ZZ$			& $35\pm3\pm6$ \\

Fake leptons		& $15\pm3\pm5$ \\

Other backgrounds	& $9.5\pm0.3\pm2.4$\\

\hline
Total background	& $392\pm7\pm19$ \\

\hline
Data			& $405$ \\
\hline
Signal efficiency $[\times 10^{-4}]$	& $9.8\pm0.1\pm1.0$ \\
\hline
\end{tabular}
\end{center}

\end{table}

The \ttz\ control region is defined by requiring at least three leptons, two 
of them with the same flavour, opposite charge and a reconstructed mass within 
15~GeV of the $Z$ boson mass. Furthermore the events are required to have at 
least two jets with $\pt > 25$~GeV and at least two $b$-tagged jets if there 
are three leptons in the event, or at least one $b$-tagged jet if there are 
four or more leptons in the event. Since the signal contribution for events 
with three leptons and two $b$-tagged jets is small, the overlap between 
signal and background regions is not removed, increasing the \ttz\ 
sensitivity in this control region. Table \ref{tab:backgrounds:ttz_yields} 
shows the yields in this control region, and the background yields agree very 
well with the data within the given uncertainty.

\begin{table}

\caption{Event yields in the \ttz\ control region for all significant sources 
of background. The first uncertainty is the statistical one associated with 
the number of events in the samples, the second uncertainty is 
systematic and is described in Section~\ref{sec:systematics}. The entry labelled ``other backgrounds'' includes all
the remaining backgrounds described in Section~\ref{sec:samples} and in 
Table~\ref{tab:samples}. The signal efficiency
is also shown.}

\label{tab:backgrounds:ttz_yields}

\begin{center}
\begin{tabular}{lc}
\hline
Sample			& Yields \\
\hline
\ttv			& $8.3\pm0.2\pm2.7$ \\
$tZ$			& $2.0\pm0.1\pm1.0$ \\
$WZ$			& $1.8\pm0.3\pm0.4$ \\

Other backgrounds	& $1.8\pm0.4\pm0.4$ \\

\hline
Total background	& $13.9\pm0.6\pm3.0$ \\
\hline
Data			& $12$ \\
\hline
Signal efficiency $[\times 10^{-4}]$	& $3.9\pm0.1\pm0.6$ \\
\hline
\end{tabular}
\end{center}

\end{table}

Backgrounds from events which contain at least one fake lepton are estimated 
from data using the matrix method~\cite{TOPQ-2011-02}. This is based on the measurement of the 
efficiencies of real and fake loose leptons to pass the nominal selection, 
$\epsilon_{\rm R}$ and $\epsilon_{\rm F}$, and on the selection of two 
orthogonal sets of events in the signal region. For the first of these sets, 
the nominal requirements are used for the leptonic selection, while for the 
second one, only the leptons which satisfy the looser selection (as described 
in Section~\ref{sec:reco}) but without meeting the nominal requirements are 
considered. For the single-lepton case, the number of 
events with one fake nominal lepton is $N^{\mathrm{nominal}}_{\rm F} = 
\left(\epsilon_{\rm F}/\epsilon_{\rm R} - \epsilon_{\rm 
F}\right)\left((\epsilon_{\rm R}-1)N_{\rm T} + \epsilon_{\rm R}N_{\rm 
L}\right)$, where $N_{\rm T}$ ($N_{\rm L}$) represents the number of selected 
events in the first (second) set defined above. The method is extrapolated to 
the three-lepton topology, with a $8\times 8$ matrix that is inverted using a 
numerical method to obtain the number of events with at least one fake 
lepton. The efficiencies for real and fake leptons are estimated as a 
function of the lepton transverse momentum by a fit of the matrix method 
results to two dedicated enriched samples of real and fake leptons: a sample 
of $Z\to \ell^+\ell^-,\ell=e,\mu$ and a same-sign dilepton sample (excluding 
same-flavour events with a reconstructed mass compatible with a $Z$ boson). 
In both samples, in order to improve the modelling of fake leptons 
originating from heavy-flavour decays, only events with at least one 
additional $b$-tagged jet are considered. The efficiency $\epsilon_{\rm R}$ 
ranges from 0.74 to 0.88 (0.80 to 0.99) and $\epsilon_{\rm F}$ from 0.010 to 
0.13 (0.035 to 0.18) for electrons (muons). The relevant uncertainties are 
calculated from the discrepancy between predicted and observed number of 
events in the control region detailed below.

A control region to test the performance of the fake-lepton estimation 
method and derive its uncertainty is defined. It requires three leptons with 
$\pt < 50$~GeV (the third one with $\pt<30$~GeV), two of them having the same 
flavour, opposite charge and a reconstructed mass within 15~GeV of the $Z$ 
boson mass, at least one $b$-tagged jet with $\pt > 35$~GeV and $\MET < 
40$~GeV. As for the \ttz\ control region, there is a small overlap with the 
signal region, which is not removed in order to increase the sensitivity to 
the fake-lepton backgrounds. The yields are shown in 
Table~\ref{tab:backgrounds:mm_cr_yields} and agree well between data and 
expectation. As a validation of the matrix method, the background in which 
exactly one of the leptons is a fake lepton is also evaluated using simulated 
samples. The results for the signal region and different control regions are 
consistent between the two methods within the estimated uncertainties.

\begin{table}

\caption{Event yields in the fake-lepton control region for all significant 
sources of background. The first uncertainty is the statistical one 
associated with the number of events in the samples, the second 
uncertainty is systematic and is described in Section~\ref{sec:systematics}. The entry labelled 
``other backgrounds'' includes all
the remaining backgrounds described in Section~\ref{sec:samples} and in 
Table~\ref{tab:samples}. The signal efficiency
is also shown.}

\label{tab:backgrounds:mm_cr_yields}

\begin{center}
\begin{tabular}{lc}
\hline
Sample & Yields\\
\hline
Fake leptons		& $7\pm1\pm4$		\\
$WZ$			& $2.7\pm0.4\pm0.7$		\\
$ZZ$			& $1.7\pm0.6\pm0.8$		\\
Other backgrounds	& $1.7\pm0.1\pm0.6$		\\
\hline
Total background	& $13\pm1\pm4$		\\
\hline
Data			& $17$				\\
\hline
Signal efficiency $[\times 10^{-4}]$	& $1.77\pm0.06\pm0.20$	\\
\hline
\end{tabular}
\end{center}

\end{table}

\begin{figure}
\begin{center}
  \includegraphics[width=.49\textwidth]{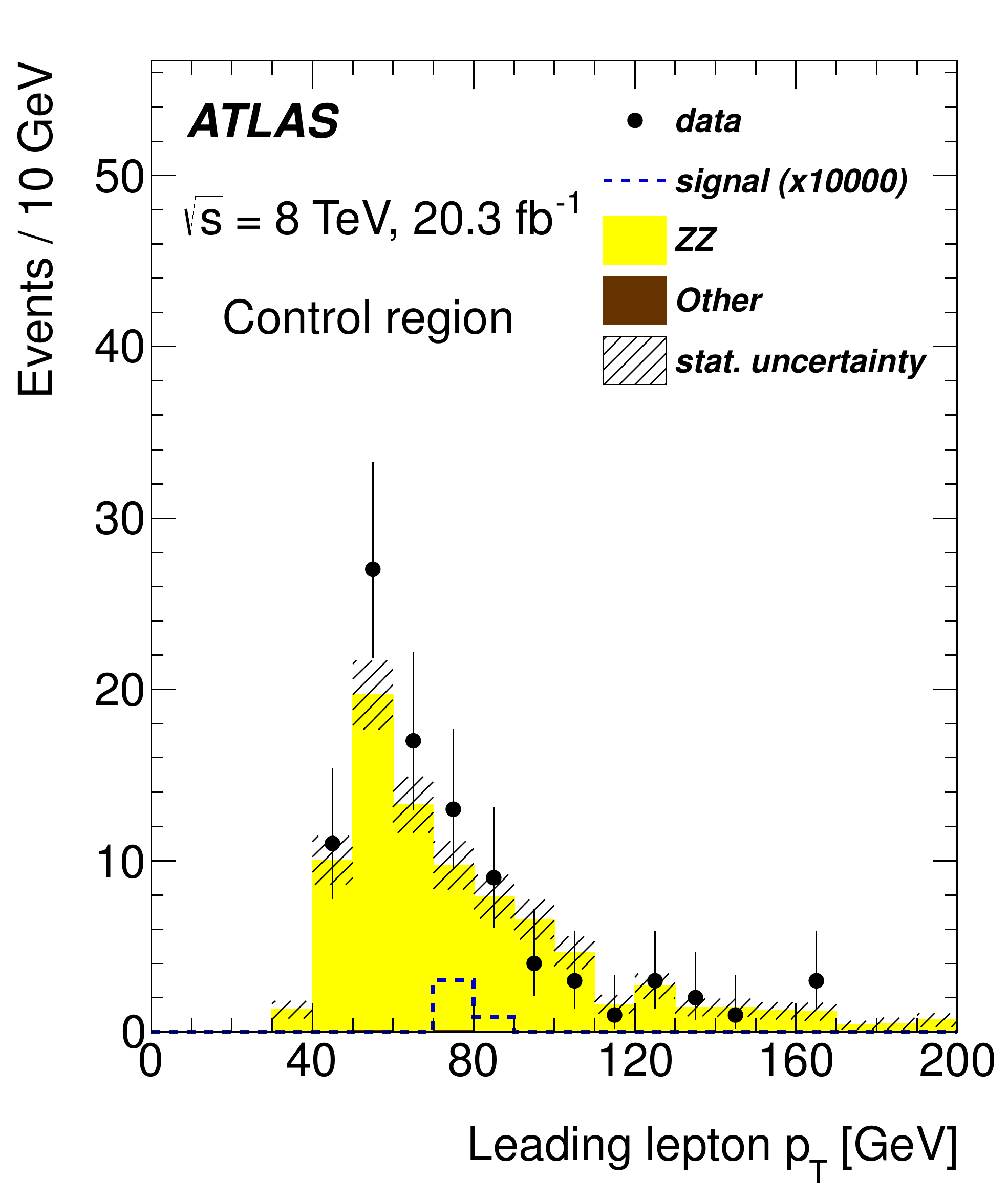}
  \includegraphics[width=.49\textwidth]{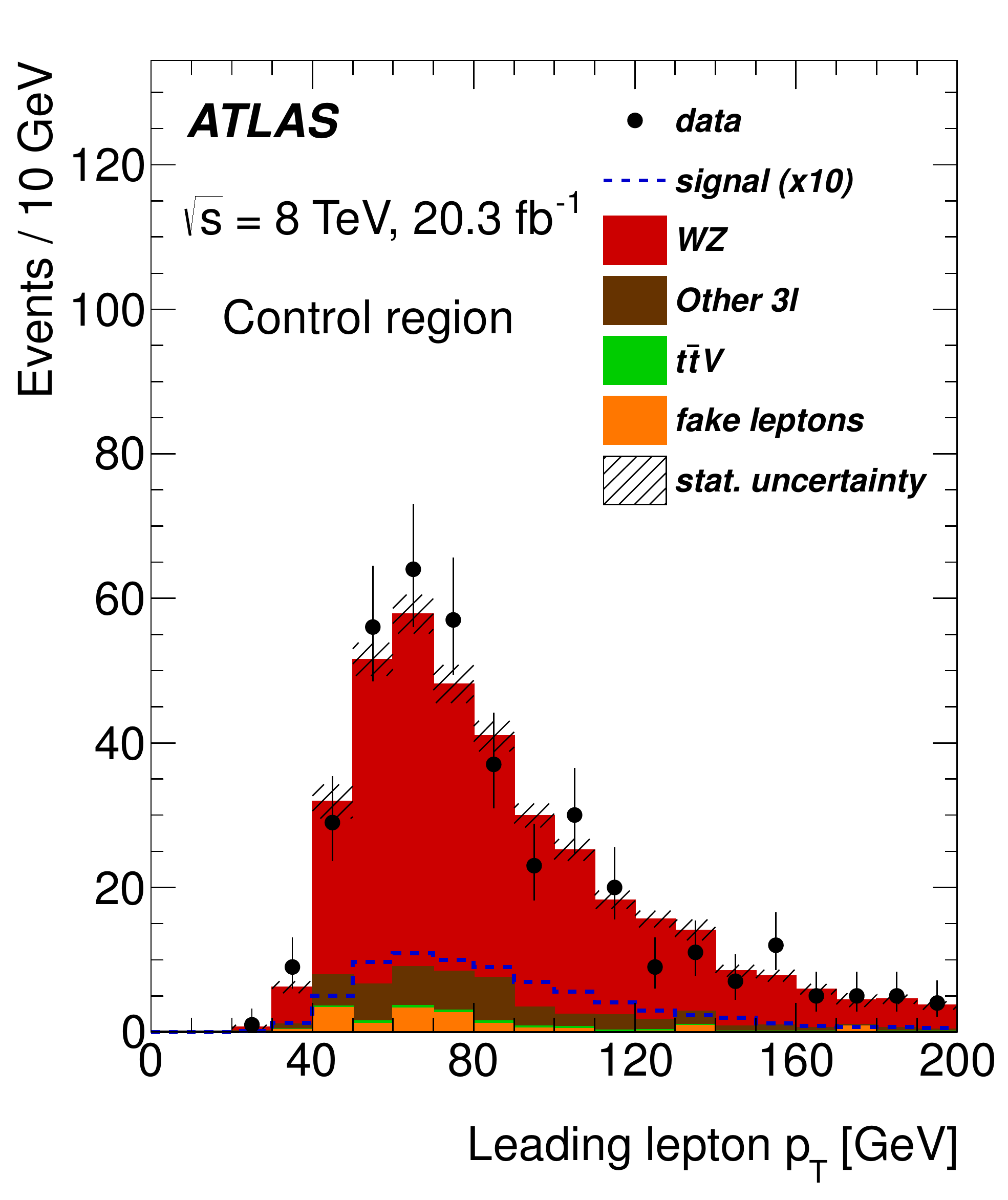}
  (a)\hspace*{.48\textwidth}(b)

  \includegraphics[width=.49\textwidth]{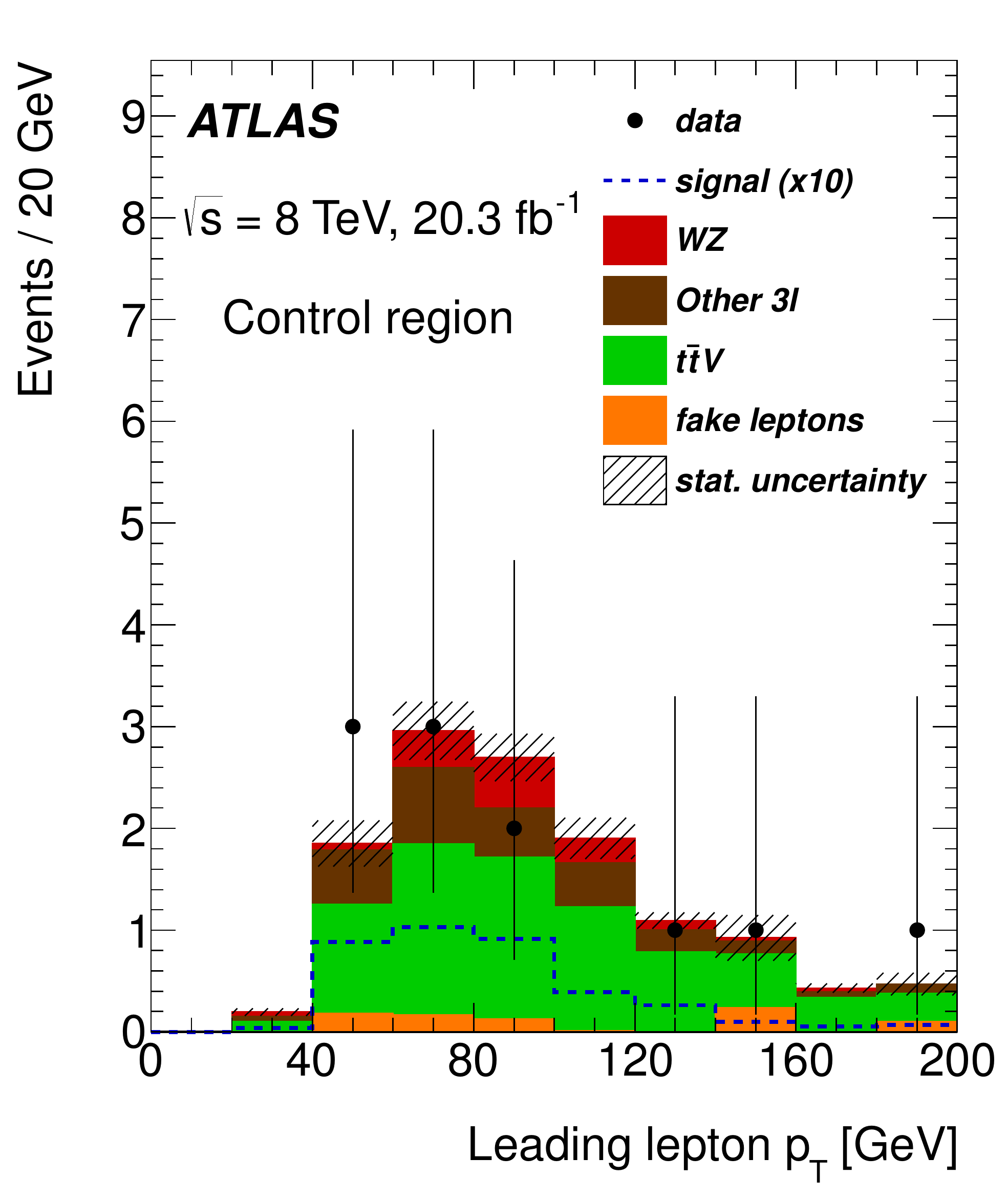}
  \includegraphics[width=.49\textwidth]{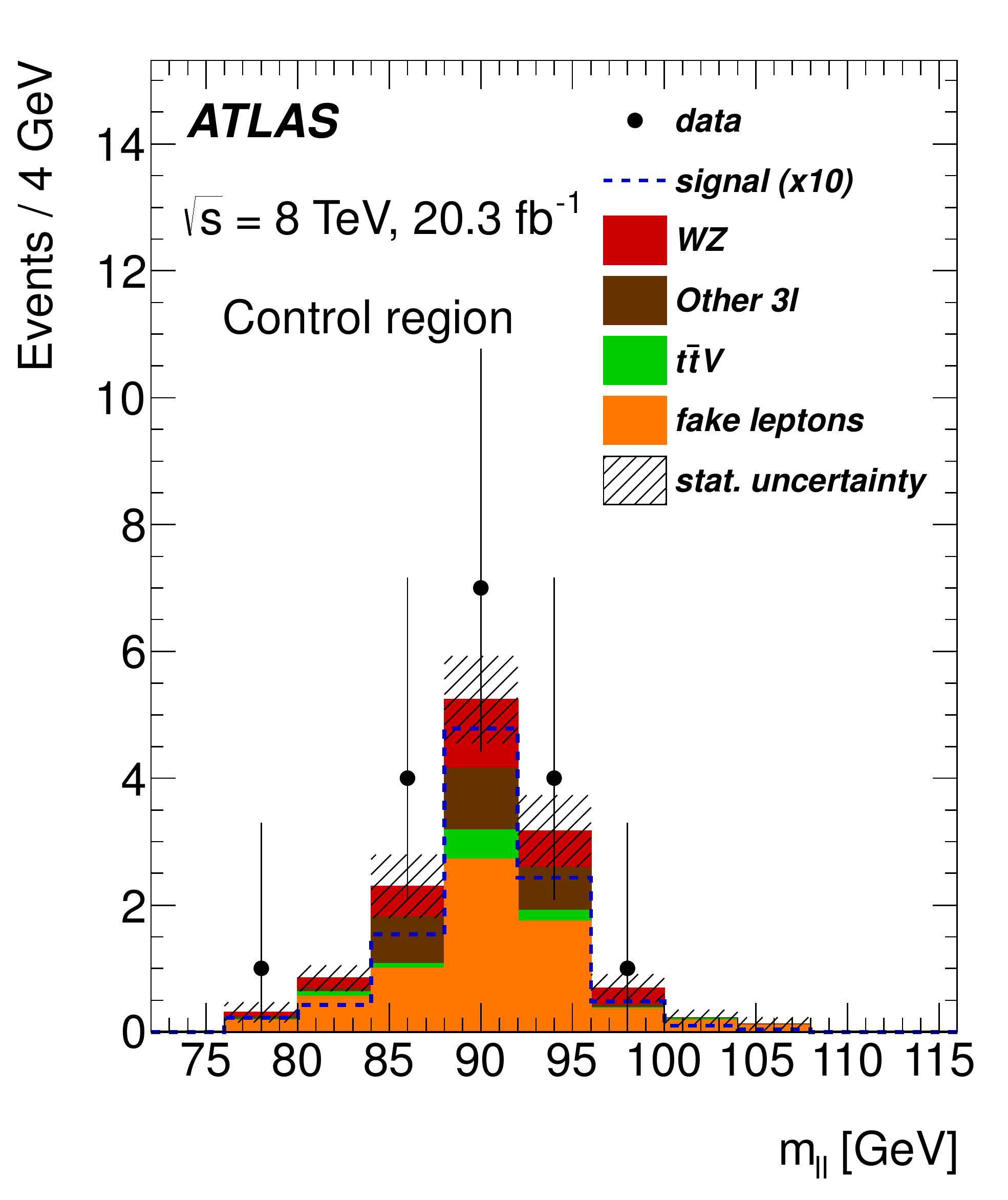}

  (c)\hspace*{.48\textwidth}(d)

\end{center}

  \caption{Expected (filled histogram) and observed (points with error bars) distributions for the \pt\ of the leading lepton in the
  (a) $ZZ$,
  (b) $WZ$ and
  (c) $t\bar tZ$ control regions and
  (d) reconstructed mass of the two leptons with the same flavour and opposite charge in the fake-lepton control region.
  For comparison, distributions for the FCNC $\ttbar \to bWqZ$ signal (dashed
  line), scaled to $10^4$ or 10 times the observed 95\% CL limit, are also shown.
  Background statistical uncertainties associated with the number of events in the samples are represented by the hatched areas.
}

  \label{fig:backgrounds}
\end{figure}

Figure~\ref{fig:backgrounds} shows the \pt\ of the leading lepton for the $ZZ$, 
$WZ$ and \ttz\ control regions, and the reconstructed mass of the two leptons 
with the same flavour and opposite charge for the fake-lepton control 
region.

%% file: systematics.tex
\section{Systematic uncertainties\label{sec:systematics}}

The effect of each source of systematic uncertainty is studied by 
independently varying the corresponding central value and propagating this through the full analysis chain. The relative 
impact of each type of systematic uncertainty on the total background and 
signal is summarised in Table~\ref{tab:systematics}.

The main uncertainty on the backgrounds comes from their modelling, which has 
the following two contributions. The
level of agreement with data of the 
reference samples is 
assessed from the combination of the Poisson uncertainty on the available 
amount of data with the statistical uncertainty of the expected background 
yields in the dedicated control regions described in 
Section~\ref{sec:backgrounds}. The uncertainties are estimated to be 6.3\%, 
12\%, 42\% and 62\%, for the $WZ$, $ZZ$, \ttz\ and fake-lepton backgrounds, 
respectively. The other contribution comes from the uncertainty on the theoretical prediction in the
signal region
and is estimated using the 
alternative $WZ$ and $ZZ$ simulated samples. The corresponding uncertainties are 17\% and 
100\%, respectively. Similarly, for \ttz, $tZ$ and Higgs samples, 
conservative values of 30\%~\cite{Campbell:2012dh,Garzelli:2012bn}, 
50\%~\cite{Campbell:2013yla} and 15\%~\cite{Heinemeyer:2013tqa} respectively, 
are used, in order to account for the theoretical uncertainties. The 
combination of all these uncertainties gives a 17\% effect on the total 
background estimation.

The theoretical uncertainties of the signal modelling, as described in 
Section~\ref{sec:samples}, namely production cross section and ISR/FSR 
modelling, are found to be 5.5\%.

For both the estimated signal and background event yields, experimental 
uncertainties resulting from detector effects are considered. The lepton 
reconstruction, identification and trigger efficiencies, as well as lepton momentum 
scales and resolutions~\cite{PERF-2013-03, PERF-2013-05, PERF-2014-05} are 
considered. The overall effect on the total background yield and the signal 
efficiency is estimated to be 4.7\% and 2.9\% respectively. The effect of the 
jet energy scale and resolution~\cite{PERF-2012-01, PERF-2011-04} 
uncertainties are evaluated as 7.7\% and 4.9\% for the background and signal, 
respectively. The $b$-tagging performance component, which includes the 
uncertainty of the $b$-, $c$-, mistagged- and $\tau$-jet scale factors (the $\tau$ 
and charm uncertainties are highly correlated and evaluated as such) is 
evaluated by varying the $\eta$-, \pt- and flavour-dependent scale factors 
applied to each jet in the simulated samples. It is estimated to be 3.9\% for 
the total background and 7.2\% for the signal efficiency. The 
$\MET$ scale uncertainty~\cite{PERF-2011-07} is found to vary the total 
background yield and the signal efficiency by 3.2\% and 1.5\%, respectively. 
All these detector systematic uncertainties are treated as fully correlated 
between signal and background.

The uncertainty related to the integrated luminosity for the dataset used in 
this analysis is 2.8\%. It is derived following the methodology described in 
Ref.~\cite{DAPR-2010-01}. It only affects the estimations obtained from 
simulated samples, therefore its impact on the total background yield 
estimation is 2.4\%.

\begin{table}

\caption{Summary of the impact of each type of uncertainty on the total 
background and signal yields. The values are shown as the relative variations 
from the nominal values. The statistical uncertainty associated with the 
number of events in the simulated samples is also shown.}

\label{tab:systematics}

\begin{center}
\begin{tabular}{lcc}
\hline
Source			& Background [\%]	& Signal [\%]	\\
\hline
Background modelling	& 17\phantom{.0}	& ---		\\
Signal modelling	& ---			& 5.5		\\
Leptons			& \phantom{0}4.7	& 2.9		\\
Jets			& \phantom{0}7.7	& 4.9		\\
$b$-tagging		& \phantom{0}3.9	& 7.2		\\
\MET			& \phantom{0}3.2	& 1.5		\\
Luminosity		& \phantom{0}2.4	& 2.8		\\
\hline
Statistical		& \phantom{0}8.1	& 1.5		\\
\hline
\end{tabular}
\end{center}

\end{table}

%% file: limits.tex
\section{Results\label{sec:limits}}

\begin{figure}
\begin{center}
  \includegraphics[width=.32\textwidth]{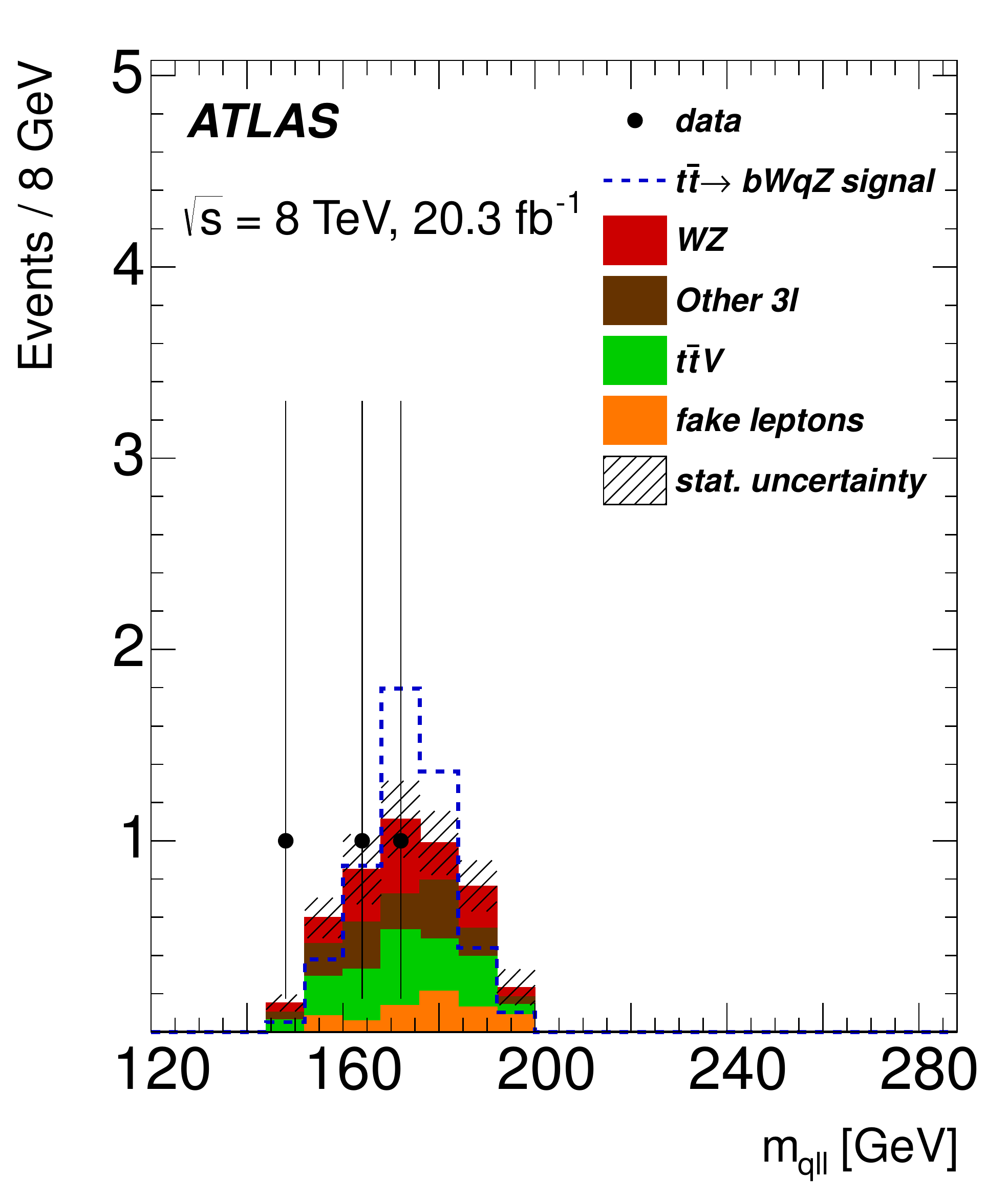}
  \includegraphics[width=.32\textwidth]{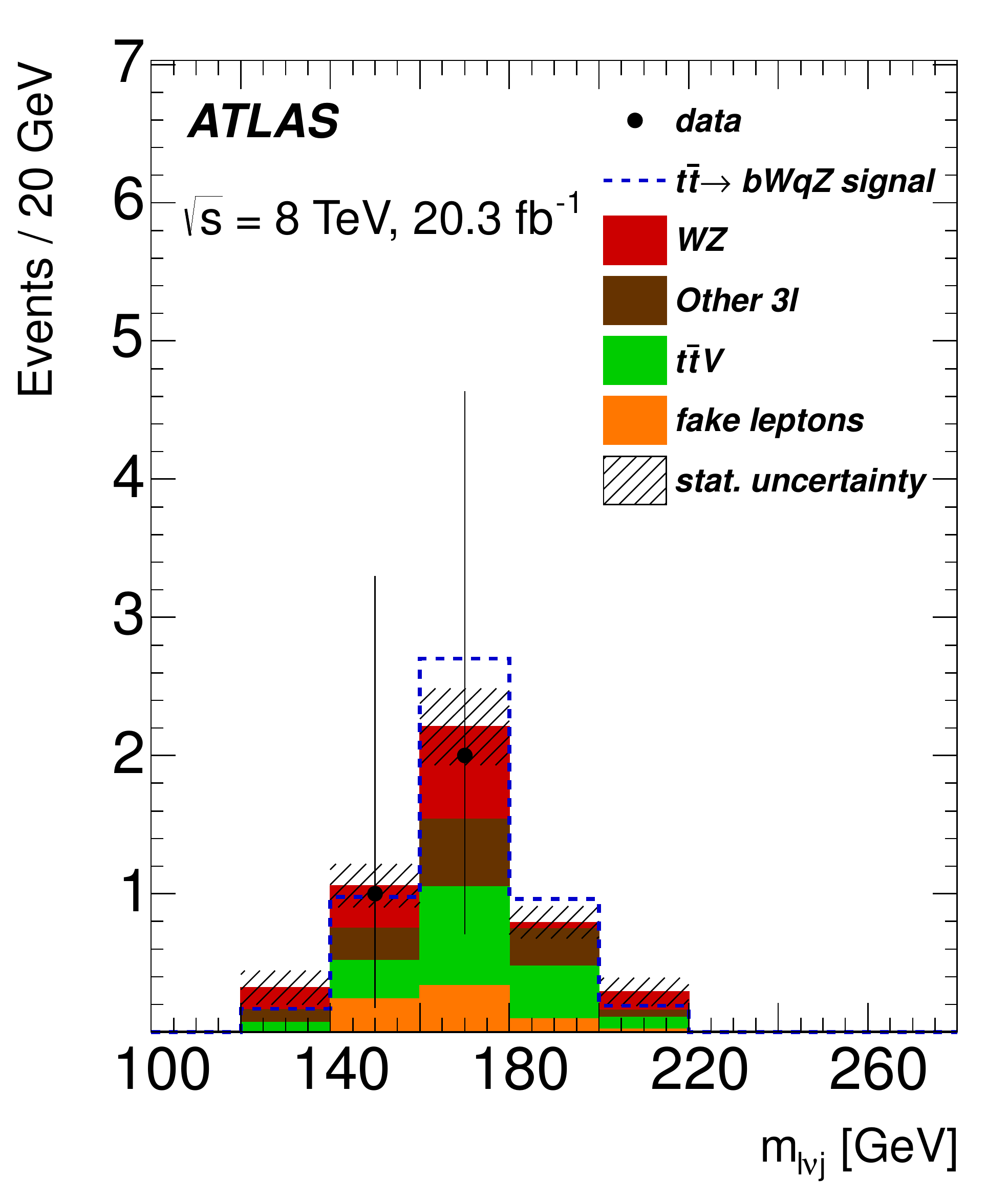}
  \includegraphics[width=.32\textwidth]{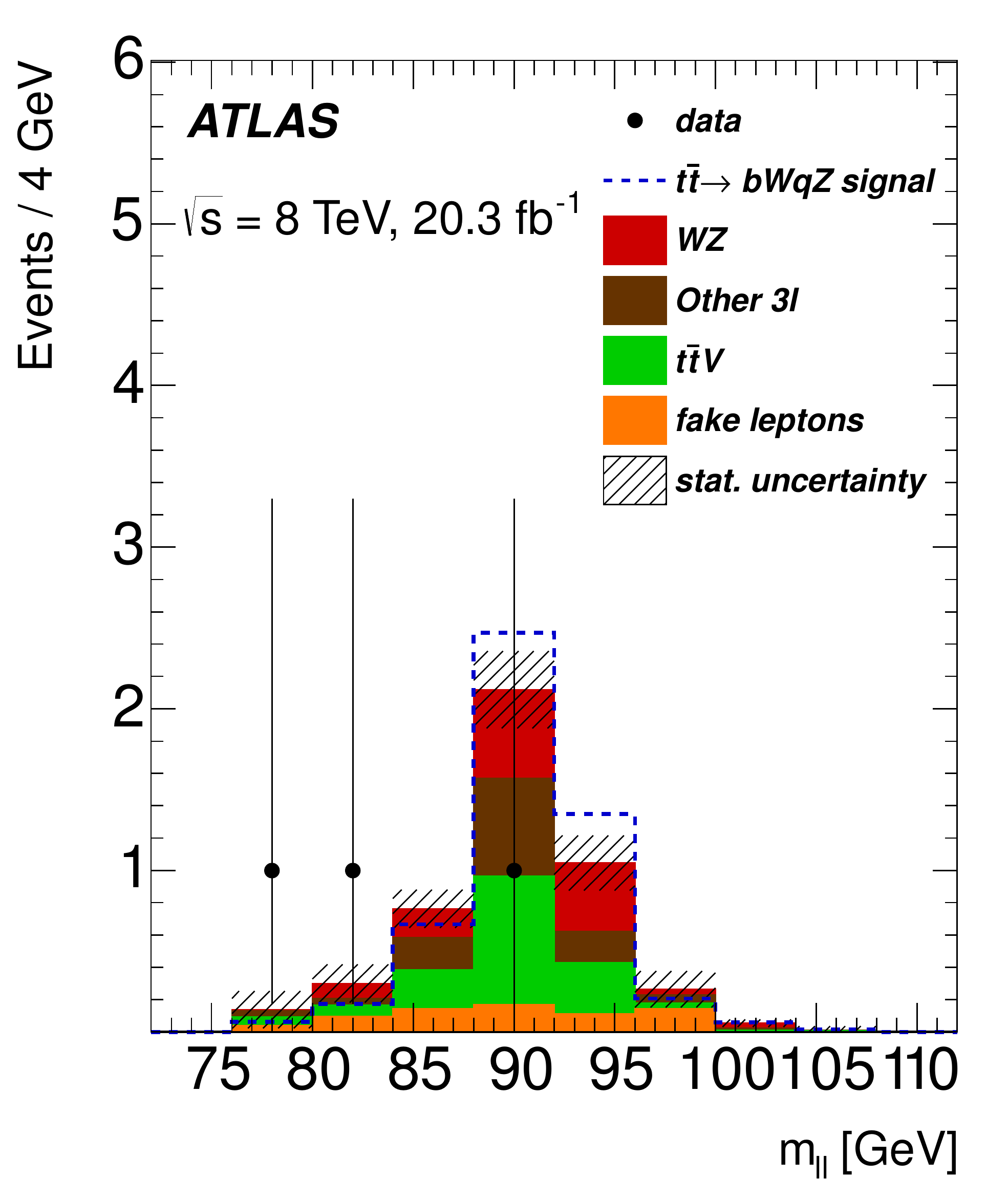}

  (a)\hspace*{.31\textwidth}(b)\hspace*{.31\textwidth}(c)

\end{center}
  \caption{Expected (filled histogram) and observed (points with error bars) distributions in the signal region after the final selection is applied for the reconstructed masses of the
  (a) top quark from the FCNC decay,
  (b) top quark from the SM decay and
  (c) $Z$ boson.
  For comparison, distributions for the FCNC $\ttbar \to bWqZ$ signal (dashed 
  line), normalised to the observed 95\% CL limit, are also shown.
  Background statistical uncertainties are represented by the hatched areas.
}

  \label{fig:selection}
  \par\vspace{0pt}
\end{figure}

Table~\ref{tab:sel7} shows the expected number of background events, number 
of selected data events and signal efficiency after the final event selection 
described in Section~\ref{sec:selection}. Figure~\ref{fig:selection} shows 
the reconstructed masses of the top quarks and $Z$ boson after the final 
selection. Good agreement between data and background yields is observed at 
all stages of the analysis. No evidence for the $t\to qZ$ decay is found and 
a 95\% CL upper limit on the number of signal events is derived using the 
modified frequentist likelihood method~\cite{Read:2002hq, Junk:1999kv}.

\begin{table}

\caption{Expected number of background events, number of selected data events 
and signal efficiency (normalised to all decays of the $W$ and $Z$ bosons), 
after the final selection. The first uncertainty is the statistical one 
associated with the number of events in the samples, the second 
uncertainty is systematic and is described in Section~\ref{sec:systematics}. The entry labelled ``other backgrounds'' includes all the remaining
backgrounds described in Section~\ref{sec:samples} and in
Table~\ref{tab:samples}.}

\label{tab:sel7}

\begin{center}
\begin{tabular}{lc}
\hline
Sample			& Yields                            \\
\hline
$WZ$			& $1.3\pm0.2\pm0.6$          \\
$t\bar tV$		& $1.5\pm0.1\pm0.5$          \\
$tZ$			& $1.0\pm0.1\pm0.5$          \\
Fake leptons		& $0.7\pm0.3\pm0.4$          \\
Other backgrounds	& $0.2\pm0.1\pm0.1$ \\
\hline
Total background        & $4.7\pm0.4\pm1.0$             \\
\hline
Data                    & 3                             \\
\hline
Signal efficiency $[\times 10^{-4}]$  & $7.8\pm0.1\pm0.8$    \\
\hline
\end{tabular}
\end{center}

\end{table}

The test-statistic $X_d$, which compares the number of observed data events 
with background and signal expectations, is defined as:
\begin{equation}
  X_d=n\ln\left(1+\frac{s}{b}\right)
  \label{fast:eq:testeestast}
\end{equation}
where $n$, $s$ and $b$ are the numbers of data, expected background and 
signal events, respectively. The $X_d$ statistical test is then compared to 
$10^5$ pseudo-experiments for the hypotheses of signal plus background 
($X_{s+b}$) and background-only ($X_b$), which are obtained by replacing $n$ 
with the corresponding number of events produced by each pseudo-experiment. The 
statistical fluctuations of the pseudo-experiments are implemented assuming 
that the number of events follows a Poisson distribution.
All statistical and systematic 
uncertainties on the expected backgrounds and signal efficiencies, as 
described in Section~\ref{sec:systematics},
are taken into account and implemented assuming Gaussian distributions.

The CL for a given signal hypothesis $s$ is defined as~\cite{Read:2002hq}:
\begin{equation}
  1-\mathrm{CL}={\int_0^{X_d}P_{s+b}(X)\mathrm{d}X \over
        \int_0^{X_d}P_{b}(X)\mathrm{d}X}\, ,
  \label{fast:eq:cl}
\end{equation}
where $P_{s+b}$ and $P_b$ are the probability density functions obtained from 
the pseudo-experiments for the $X_{s+b}$ and $X_b$ values, respectively, and 
are functions of $s$ and $b$. The limit on the number of signal events is 
determined by finding the value of $s$ corresponding to a CL of 95\%. 
The expected limit is computed by replacing $X_d$ with the median of 
the statistical test for the background hypothesis ($X_b$).

The limits on the number of signal events are converted into upper limits on 
the $t\to qZ$ branching fraction using the NNLO+NNLL calculation, and 
uncertainty, for the $t\bar t$ cross section, and constraining BR$(t\to 
bW)=1-\mathrm{BR}(t\to qZ)$. Table~\ref{tab:limits} shows the observed limit on 
BR$(t\to qZ)$ together with the expected limit and corresponding $\pm1\sigma$ 
bounds. These values are calculated using the reference $t\bar t\to bWcZ$ 
sample,
since it gives a more conservative result than the $t\bar t\to bWuZ$
sample.
The smaller $b$-tagged jet multiplicity of the $t\bar t\to bWuZ$ signal 
sample leads to an improvement of 4\% in the limit.

\begin{table}
\caption[95\% CL limits]{Observed and expected 95\% CL limits on the FCNC 
top-quark decay BRs. The expected 
central value is shown together with the $\pm1\sigma$ bands, which 
includes the contribution from the statistical and systematic uncertainties.}
\label{tab:limits}

\begin{center}
\begin{tabular}{lr}
\hline
Observed	& $7\times 10^{-4}$ \\
\hline
$(-1\sigma)$	& $6\times 10^{-4}$ \\
Expected	& $8\times 10^{-4}$ \\
$(+1\sigma)$	& $12\times 10^{-4}$\\
\hline
\end{tabular}
\end{center}

\end{table}

\begin{figure}

\begin{center}
  \includegraphics[width=.49\textwidth]{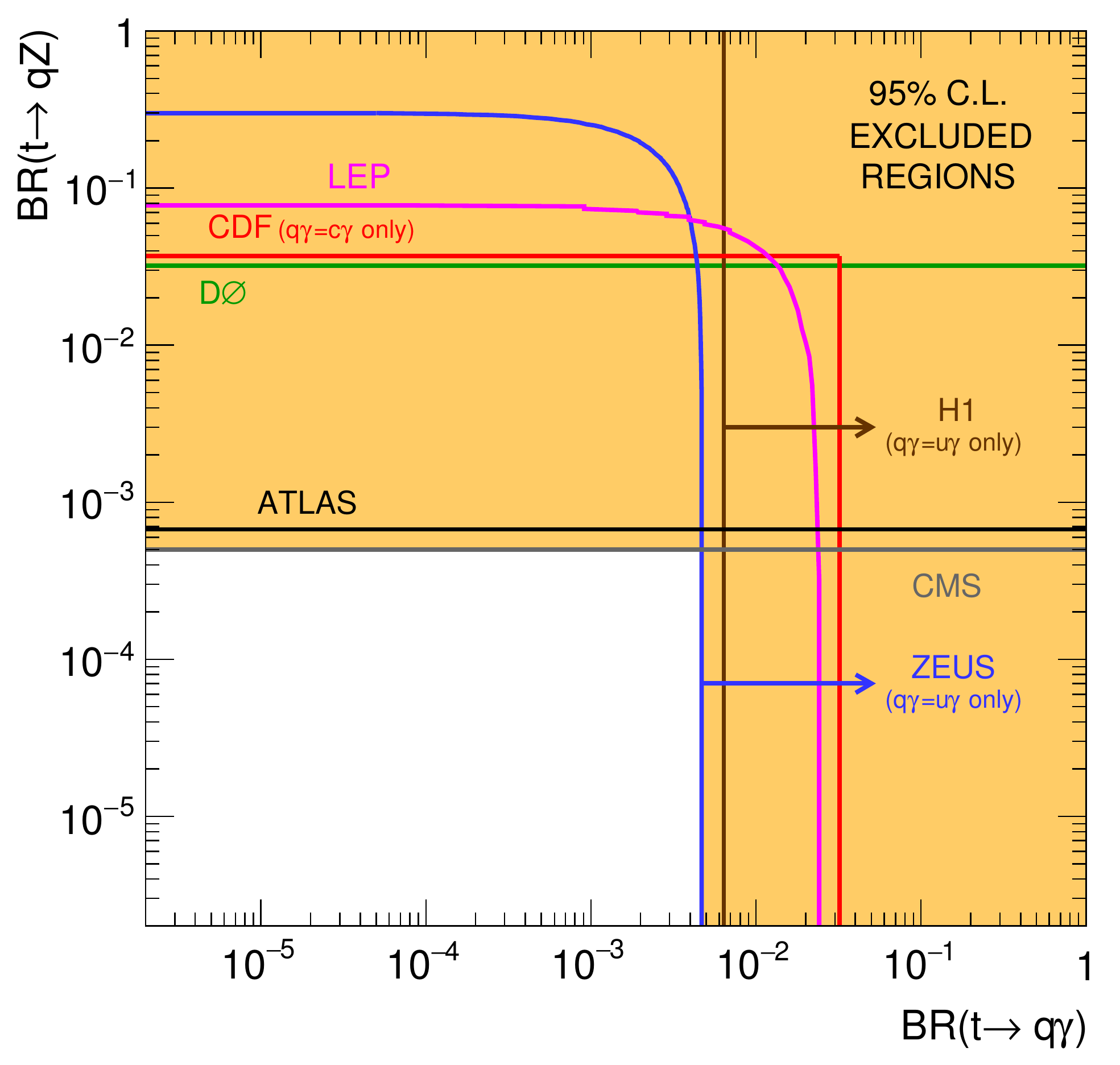} 
  \includegraphics[width=.49\textwidth]{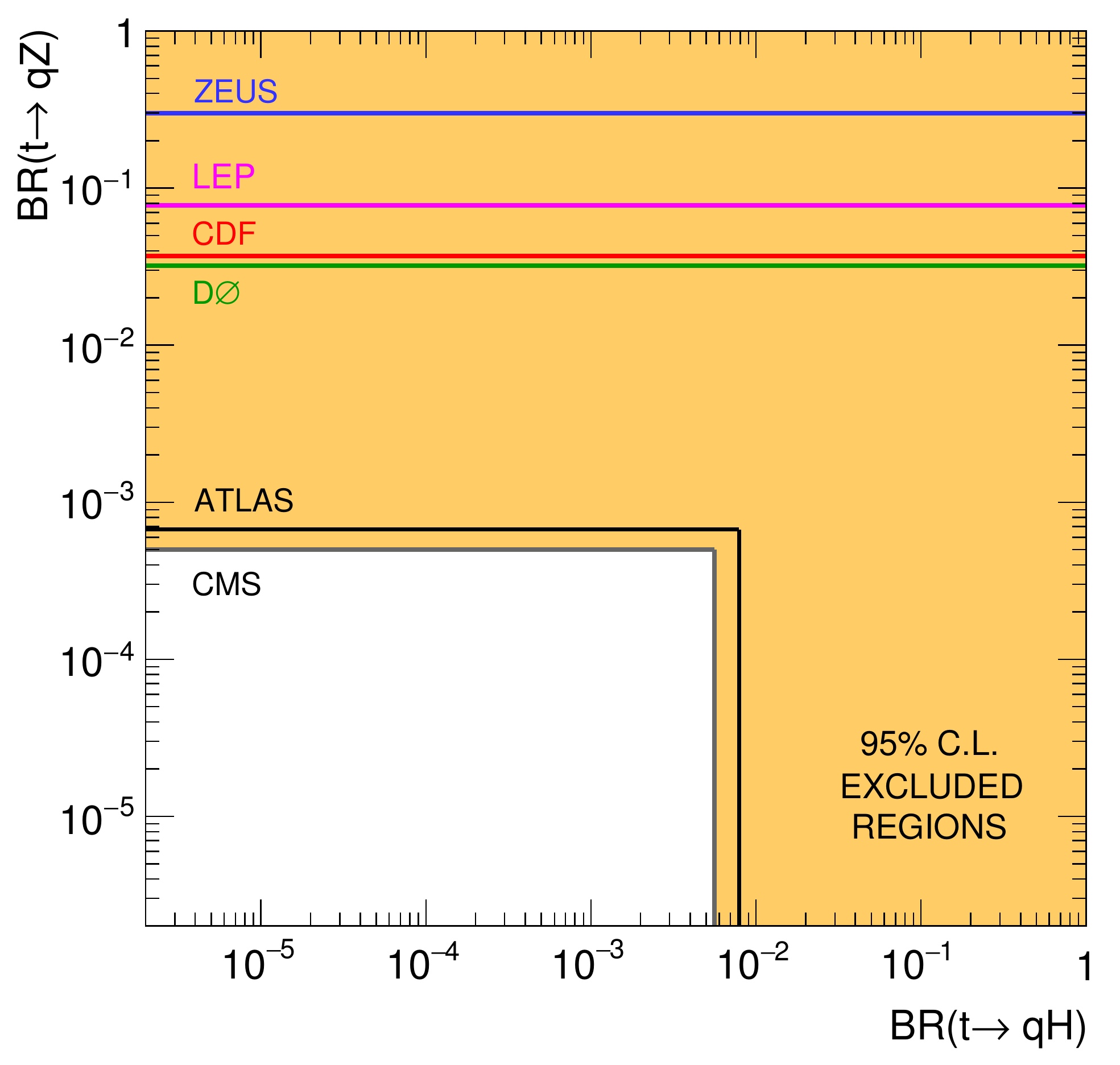} 

  (a)\hspace*{.48\textwidth}(b)
\end{center}

\caption{The current 95\% CL observed limits on the (a) BR$(t\to q\gamma)$ vs 
BR$(t\to qZ)$ and (b) BR$(t\to qH)$ vs
BR$(t\to qZ)$ planes are shown. The full lines represent the results from the
ATLAS~\cite{HIGG-2013-09},
CDF~\cite{Aaltonen:2008ac, Abe:1997fz}
CMS~\cite{CMS-TOP-12-037, CMS-HIG-13-025}, 
D\O~\cite{Abazov:2011qf},
H1~\cite{Aaron:2009vv},
LEP (combined results of the ALEPH, DELPHI, L3 and OPAL collaborations)~\cite{Heister:2002xv, Abdallah:2003wf, Abbiendi:2001wk, Achard:2002vv, LEP-Exotica-WG-2001-01}
and 
ZEUS~\cite{Abramowicz:2011tv} collaborations.
The ATLAS lines correspond to the 
limit on BR($t\to qZ$) set in this paper.}

\label{fig:brplane}
\end{figure}

Figure~\ref{fig:brplane} compares the 95\% CL observed limit found in this analysis
with the results
from other FCNC searches performed by the H1, ZEUS, LEP (combined results of the ALEPH, 
DELPHI, L3 and OPAL collaborations), CDF, D\O\ and CMS collaborations. The 
results presented in this paper are consistent with the ones from the CMS 
Collaboration.

%% file: conclusions.tex
\FloatBarrier
\section{Conclusions\label{sec:conclusions}}

A search for the FCNC top-quark decay $t\to qZ$ in events with three leptons 
has been performed using LHC data collected by the ATLAS experiment at a 
centre-of-mass energy of $\sqrt{s} = 8$~TeV and corresponding to an 
integrated luminosity of 20.3 fb$^{-1}$ recorded in 2012. No evidence for 
signal events is found and a 95\% CL limit for the $t\to qZ$ branching 
fraction is established at BR$(t\to qZ) < 7\times 10^{-4}$, in agreement with 
the expected limit of BR$(t\to qZ) < 8\times 10^{-4}$.

%% file: Acknowledgements.tex

We thank CERN for the very successful operation of the LHC, as well as the
support staff from our institutions without whom ATLAS could not be
operated efficiently.

We acknowledge the support of ANPCyT, Argentina; YerPhI, Armenia; ARC, Australia; BMWFW and FWF, Austria; ANAS, Azerbaijan; SSTC, Belarus; CNPq and FAPESP, Brazil; NSERC, NRC and CFI, Canada; CERN; CONICYT, Chile; CAS, MOST and NSFC, China; COLCIENCIAS, Colombia; MSMT CR, MPO CR and VSC CR, Czech Republic; DNRF, DNSRC and Lundbeck Foundation, Denmark; IN2P3-CNRS, CEA-DSM/IRFU, France; GNSF, Georgia; BMBF, HGF, and MPG, Germany; GSRT, Greece; RGC, Hong Kong SAR, China; ISF, I-CORE and Benoziyo Center, Israel; INFN, Italy; MEXT and JSPS, Japan; CNRST, Morocco; FOM and NWO, Netherlands; RCN, Norway; MNiSW and NCN, Poland; FCT, Portugal; MNE/IFA, Romania; MES of Russia and NRC KI, Russian Federation; JINR; MESTD, Serbia; MSSR, Slovakia; ARRS and MIZ\v{S}, Slovenia; DST/NRF, South Africa; MINECO, Spain; SRC and Wallenberg Foundation, Sweden; SERI, SNSF and Cantons of Bern and Geneva, Switzerland; MOST, Taiwan; TAEK, Turkey; STFC, United Kingdom; DOE and NSF, United States of America. In addition, individual groups and members have received support from BCKDF, the Canada Council, CANARIE, CRC, Compute Canada, FQRNT, and the Ontario Innovation Trust, Canada; EPLANET, ERC, FP7, Horizon 2020 and Marie Skłodowska-Curie Actions, European Union; Investissements d'Avenir Labex and Idex, ANR, Region Auvergne and Fondation Partager le Savoir, France; DFG and AvH Foundation, Germany; Herakleitos, Thales and Aristeia programmes co-financed by EU-ESF and the Greek NSRF; BSF, GIF and Minerva, Israel; BRF, Norway; the Royal Society and Leverhulme Trust, United Kingdom.

The crucial computing support from all WLCG partners is acknowledged
gratefully, in particular from CERN and the ATLAS Tier-1 facilities at
TRIUMF (Canada), NDGF (Denmark, Norway, Sweden), CC-IN2P3 (France),
KIT/GridKA (Germany), INFN-CNAF (Italy), NL-T1 (Netherlands), PIC (Spain),
ASGC (Taiwan), RAL (UK) and BNL (USA) and in the Tier-2 facilities
worldwide.

%% file: atlas_authlist.tex
\begin{flushleft}
{\Large The ATLAS Collaboration}

\bigskip

G.~Aad$^{\rm 85}$,
B.~Abbott$^{\rm 113}$,
J.~Abdallah$^{\rm 151}$,
O.~Abdinov$^{\rm 11}$,
R.~Aben$^{\rm 107}$,
M.~Abolins$^{\rm 90}$,
O.S.~AbouZeid$^{\rm 158}$,
H.~Abramowicz$^{\rm 153}$,
H.~Abreu$^{\rm 152}$,
R.~Abreu$^{\rm 116}$
% ...
\bigskip
\\
% ...
$^{11}$ Institute of Physics, Azerbaijan Academy of Sciences, Baku, Azerbaijan\\
% ...
$^{85}$ CPPM, Aix-Marseille Universit{\'e} and CNRS/IN2P3, Marseille, France\\
% ...
$^{90}$ Department of Physics and Astronomy, Michigan State University, East Lansing MI, United States of America\\
% ...
$^{107}$ Nikhef National Institute for Subatomic Physics and University of Amsterdam, Amsterdam, Netherlands\\
% ...
$^{113}$ Homer L. Dodge Department of Physics and Astronomy, University of Oklahoma, Norman OK, United States of America\\
% ...
$^{116}$ Center for High Energy Physics, University of Oregon, Eugene OR, United States of America\\
% ...
$^{151}$ Institute of Physics, Academia Sinica, Taipei, Taiwan\\
$^{152}$ Department of Physics, Technion: Israel Institute of Technology, Haifa, Israel\\
$^{153}$ Raymond and Beverly Sackler School of Physics and Astronomy, Tel Aviv University, Tel Aviv, Israel\\
% ...
$^{158}$ Department of Physics, University of Toronto, Toronto ON, Canada\\
% ...
\end{flushleft}
